\documentclass[showpacs,aps,preprintnumbers,prb,amsmath,amssymb]{revtex4}
\usepackage{graphicx}

\begin{document}
\preprint{APS/123-QED}
\title{Stabilization of the collective Kondo semiconducting state by Sb doping
in CeNiSn$_{1-x}$Sb$_{x}$ and the criterion of its appearance}

\author{Jozef Spa\l{}ek}
\email{ufspalek@if.uj.edu.pl}
\homepage{http://th-www.if.uj.edu.pl/ztms/jspalek_een.htm}
\affiliation{Marian Smoluchowski Institute of Physics, Jagiellonian 
University, Reymonta 4, 30-059 Krak\'ow, Poland }

\author{Andrzej \'{S}lebarski}
\email{slebar@us.edu.pl}
\affiliation{Institute of Physics, University of Silesia, Uniwersytecka 4, 
40-007 Katowice, Poland }
\date{\today}
\begin{abstract}

{\em Semimetallic\/} CeNiSn is shown to transform into a {\em Kondo semiconductor\/} upon the substitution of few percent of Sb
for Sn. The full-gap formation is not decisively influenced by the atomic
disorder introduced by the substitution. Instead, the extra valence electrons
introduced with the Sb doping (one per Sb 
atom) 
contribute to the formation of the collective {\em Kondo spin-singlet\/} state 
at low temperatures, as seen by a reduction of magnetic susceptibility. 
The definition of the Kondo semiconductor is provided and
the difference with either 
the band-Kondo or the Mott-Hubbard insulators is stressed.

\end{abstract}

\maketitle

Kondo insulators (KI) 
have been discovered some time ago [1] and 
belong to the class of either nonmagnetic semiconductors with the narrowest gap 
known or to semimetals, both with a heavy-fermion metallic state setting in
gradually at elevated 
temperatures $T>\Delta$. 
Their nature is known to a lesser extent than that of heavy fermion metals. 
CeRhSb is an example of the full-gap semiconductor with the conductivity gap 
$\Delta\simeq\,7.6$~K,\cite{1} whereas CeNiSn is a semimetallic system\cite{2}
with the
gap vanishing at least in 
some directions in reciprocal space. Recently, we have discovered\cite{3} a quantum critical point for the system CeRhSb$_{1-
x}$Sn$_{x}$ with $x\sim 0.12$, when the system undergoes a transition from the KI state to the metallic ({\em non-Fermi 
liquid\/}, NFL) state. In that system the carrier concentration diminishes upon the Sn substitution 
for Sb. Therefore, one can ask the basic questions: what happens when by doping we act in the opposite direction, i.e. increase the carrier concentration, 
(e.g. synthetize the system CeNiSn$_{1-x}$Sb$_{x}$)? Will the increased concentration of carriers 
produce a Kondo semiconducting state with a Kondo-singlet collective state reducing essentially the 
magnetic susceptibility and thus producing at the same time a gap, which can be seen in the temperature 
dependence of the electrical resistivity $\rho(T)$? 
We show here that this is indeed the case, i.e. the substitution of $2\%$ of Sb 
for Sn in CeNiSn produces a gap of magnitude $\Delta\sim 4$~K
(the polycrystalline-sample data for $x=0$ exhibit 
an overall gap $\Delta\simeq 1.7$~K). 
The gap, associated in the following with the {\em collective-Kondo-state 
formation\/}, is most directly singled out by an intrinsic magnetic 
susceptibility $\chi(T)$ reduction when lowering the temperature. 
Furthermore, the 
universal scaling law $\rho(T)\chi(T)=const$ 
is obeyed when lowering $T$ and is claimed to represent a universal characteristic of
those
strongly correlated systems.
The above three features (activated behavior of $\rho(T)$, reduction of 
$\chi(T)$, $\rho(T)\chi(T)=const$)
constitute {\em unique\/}, in our view for the first time, 
complete definition 
of the Kondo semiconductor (insulator at $T=0$) from an experimental side.
In this respect, the present data confirm the earlier results for
CeRhSb$_{1-x}$Sn$_{x}$.
\cite{3}
Additionally, with the increasing temperature the system evolves 
from {\em 
Kondo semiconducting state\/} into a (moderately) heavy fermion state. We also 
provide the analysis of the systematic evolution of the CeNiSn$_{1-x}$Sb$_{x}$ systems 
as a function of $x\ll 1$. 

It follows from our analysis, that the formation of the 
Kondo-insulator state is mainly due to the formation of a {\em collective 
spin-singlet state\/} 
and {\em may not be\/} necessarily connected with the particular (integer) 
number of electrons involved. Hence in this case, the KI state is not either
a full-band insulator composed of heavy quasiparticles\cite{4}
or the Mott-Hubbard-type magnetic semiconductor, 
as the band filling is noninteger. 
The difference between the Kondo-band and the Kondo collective state
has been shown in Fig. 5 of Ref. 3b. In this paper the latter picture 
is confirmed directly.
The effect of the atomic disorder 
introduced by the substitution is not regarded as crucial (see below).

We start with presentation of our experimental results and only then draw
the conclusions specified already above. 
Polycrystalline samples of CeNiSn$_{1-x}$Sb$_{x}$ have been prepared by arc 
melting of $(1-x)$ (CeNiSn) and $x$ (CeNiSb) on a water cooled copper hearth in 
a high purity argon atmosphere with Al getter. Each sample was remelted several 
times and annealed at $800^{o}$C for 2 weeks. Analysis of the $x$-ray diffraction pattern with the Powdered-Cell program 
revealed that the samples with $x\leq 0.22$ crystallize in an orthorhombic 
$\epsilon$-TiNiSn structure. The lattice parameters of the components are 
practically $x$ independent because of similar atomic radii of Sn and Sb. 
However, measurements of the specific heat displayed weak anomalies at $\sim 
6$~K, which are attributed to the magnetic ordering of the impurity phases of 
Ce$_{2}$O$_{3}$. The impurity inclusion is present in most of the CeNiSn samples 
(see e.g. Ref.~\onlinecite{5}). Electrical resistivity measurements were carried out using a 
standard four-wire technique. The ac susceptibility was
measured in the 
magnetic field of $10$~Oe using Lake Shore susceptometer.

\begin{figure}
\includegraphics[width=0.5\textwidth]{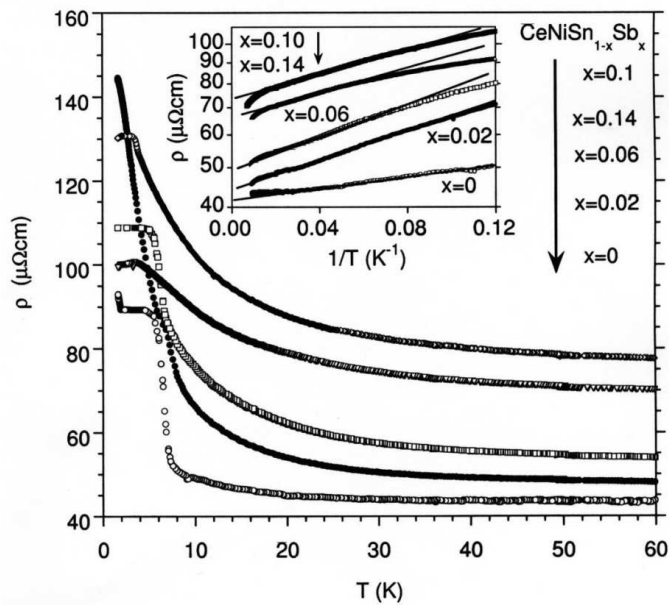}
\caption{Temperature dependence of the resistivity in linear and logarithmic (see 
inset) scales. The presence of plateau indicates the extrinsic (impurity-band) 
contribution. The activation energy is weakly dependent on $x$ for $x>0$.
}
\end{figure}

In Fig.~1 we plot the temperature dependence of the resistivity $\rho(T)$
for the CeNiSn$_{1-
x}$Sb$_{x}$ samples. Note that the substitution of Sb for Sn 
increases nominally the number of carriers by one per formula. Nevertheless, this produces an enhanced activated 
behavior $\rho(T)=\rho_{0}+A\exp(\Delta/k_{B}T)$, with the conductivity gap $\Delta\simeq 4.3\pm 
0.1$~K for $x$ in the range $0.02\div 0.06$, as compared to the value of the overall gap $\Delta\simeq 
1.7\pm 0.1$~K for the pure system NiCeSn (see the inset). 
A levelling of the resistivity is most probably due 
to the impurity band (as exemplified also by the presence of $\rho_{0}$). The circumstance that the resistivity is of the order
$10^{2}\mu\Omega$cm speaks in favor of good quality of our polycrystalline samples. The most important fact here is that 
the intentional impurities (Sb) stabilize the semiconducting-gap state and thus produce a true Kondo semiconducting state even though the doping nominally 
composes a fractional filling of the valence band.
The reference level $\rho_{0}$ of $\rho(T)$ increases with the increasing 
substitution $x$ and thus must be due to the atomic disorder.

An important conclusion can be drawn already at this stage. Namely, 
the presence of this semiconducting state cannot be related to any particular 
band filling, as would be the case for either the nonmagnetic Kondo-band\cite{4} or 
the Mott-Hubbard 
magnetic semiconductors. So, by Kondo semiconducting state we mean a binding
of all electrons involved into a
collective spin-singlet state, as discussed next.

In Fig. 2 we present the temperature dependence of $ac$ magnetic susceptibility $\chi$. While 
$\chi(T)$ for the pure system $(x=0)$ is increasing with the decreasing temperature for $T\leq 25$~K, the corresponding data for $x>0$ exhibit 
the $\chi(T)$ downturn when $T$ is lowered. The systems exhibit also a sharp 
upturn at low temperatures $T\leq 5$~K. Important is to note that in that low-$T$ range
$\chi(T)=\chi_{0}+nC/(T-\Theta)$ for pure system, with 
$\chi_{0}=5.6\cdot 10^{-4}$emu/mol, $C=1.86\cdot 10^{-3}$emu/K~mol, and 
$\Theta=-0.7$~K. In view 
of the circumstance, that the molar Curie constant for Ce$^{3+}$ ion is 
$C=0.807$~emu/K~mol, this upturn 
is ascribed to the Ce interstitial impurities of the concentration $n=0.4\%$.\cite{3,5} 
Conversely, the samples with 
$x>0$ exhibit a pronounced downturn, which is attributed to the formation of the 
collective Kondo singlet type of state. 
Although the decrease in $\chi(T)$ here is not as spectacular as in the 
CeRhSb$_{1-x}$Sn$_{x}$ case,\cite{3} it must be associated with the formation of a 
nonmagnetic collective state and the effect is the strongest for $T\leq 20$~K.
The sharp increase of $\chi$ as $T\rightarrow 0$ is regarded again as due to the 
impurity phase Ce$_{2}$O$_{3}$, as all the curves follow then the 
$x=0$ dependence, with only moderately changed parameters. Note also that the $\chi$ decrease for CeNiSn is 
almost linear with $T$ reflecting the intricasies of the pseudogap shape of the quasiparticle density of states. 
Namely, the number of excited carriers in the zero-gap sample is $\sim k_{B}T\rho(\epsilon_{F})$, where 
$\rho(\epsilon_{F})$ is the density of states at the Fermi level. Also, the magnetic susceptibility is 
proportional to that carrier concentration, if their appearance comes from breaking of the collective Kondo-spin-singlet state (with $\chi\approx 0$). 
In effect, the temperature dependence of $\chi$ will be roughly linear in $T$, 
as observed.
This is not the case in the full-gap state for $x>0$. The value of $\chi_{0}$ depends in a nonsystematic 
manner on $x$; hence, it must be ascribed to the impurity narrow-band contribution of the 
magnitude $10^{-4}-10^{-3}$~emu/mol, which is a quite large number. This means that the impurity-band electrons are quite heavy as well.

\begin{figure}
\includegraphics[width=0.5\textwidth]{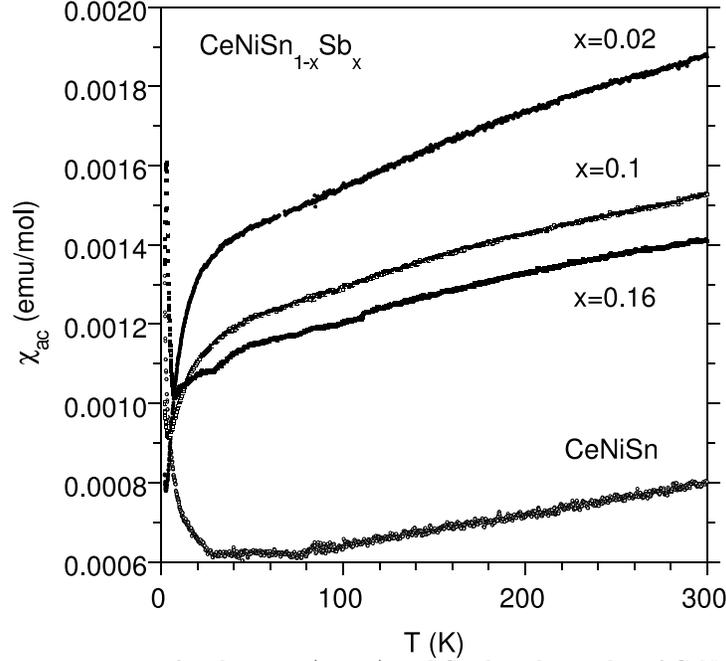}
\vspace{8mm}
\caption{The susceptibility vs. temperature for the pure $(x=0)$ and Sn doped 
samples of CeNiSb$_{1-x}$Sn$_{x}$. Note a pronounced linear $T$ behavior for 
$(x=0)$ and a value $\chi_{0}$ which is of a nonsystematic nature 
(and attributed to extrinsic 
impurity-band contribution).
}
\end{figure}

In Fig. 3 we plot the scaling $\rho^{-1}$ vs. $\chi$ and 
regard it as one of the basic properties of the Kondo semiconducting systems. This type of 
scaling is absent for the pure CeNiSn system. Therefore, only the system with 
$x>0$ can be regarded as such. 
One sees that the $\chi(T)$ diminution is a clearer sign of the onset of the
collective Kondo-singlet state rather 
than the activated character resistivity $\rho=\rho(T)$. This is partly because
the impurity band is present even for pure samples (cf. Fig. 1).
The overall decrease of $\chi$ with the increasing $x$ may be influenced 
to some degree to the disorder, but this cannot be singled out at present.

In principle, it is possible that the disorder creates localized 
states in the pseudogap region. 
This is in 
agreement with the trend observed in Fig. 1, where the most heavily doped 
samples ($x=0.1$ and $0.14$) have the highest resistivity caused by the
$\rho_{0}$ increase. 
Nevertheless, the activated behavior is reduced then ($\Delta\simeq 3.7$~K),
so the effect of the disorder should not be crucial.
Therefore, we assign 
the full-gap KI state for low $x$ as primarily 
due to the increased binding energy caused 
by the carrier-concentration increase achieved by the substitution.

\begin{figure}
\includegraphics[width=0.5\textwidth]{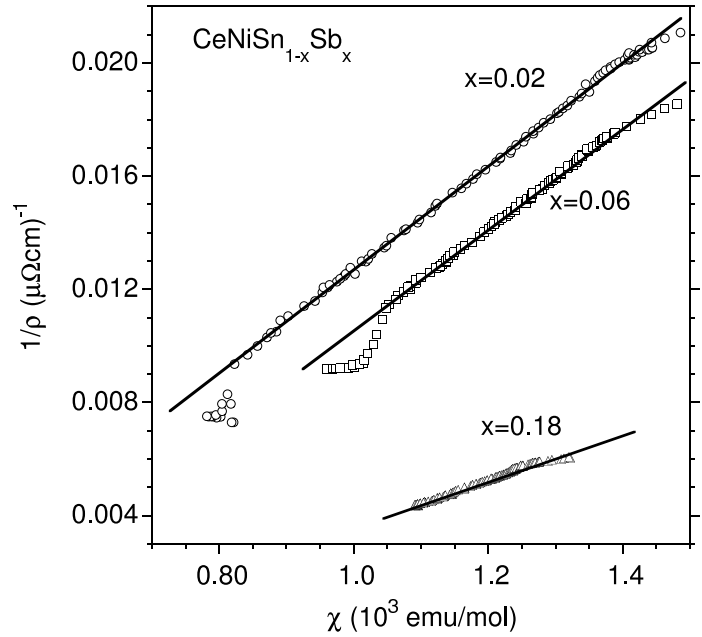}
\caption{Scaling of the inverse resistivity with the magnetic
susceptibility. The data for 
$(x=0.02)$ and $(x=0.06)$ fall into a single curve when shifted vertically.}
\end{figure}

The question remains whether the growing value of $\chi$ in the low-$T$ range 
(cf. Fig. 2) may indicate an onset of a magnetic ordering or
of enhanced magnetic correlations below 
about $5$~K. For that purpose the specific heat has been measured, as displayed 
in Fig. 4 for the sample with $x=0.18$. The value of $\gamma=174$~mJ/molK$^{2}$ 
has been extracted from the $C/T$ data vs. $T^{2}$ in the higher-$T$ regime, as 
shown. Those data indeed show for lower $x$ a cusp-like behavior near the lowest temperature 
measured. It is tempting to say that it provides an evidence for a disordered 
antiferromagnetic arrangement $(\Theta<0)$ of impurity phase containing Ce$^{3+}$ $4f$ spins. This means that the Kondo 
screening of the $(17+x)$ valence electrons of the $4f^{1}$ spin moment of 
Ce$^{3+}$ impurities 
ions is not complete but, the Kondo binding energy $(2\Delta)$ is a well defined bulk effect. 
In other words, impurity Ce$^{3+}$ spins do not hybridize with valence states.
These impurities are unfortunately always present in this class of compounds, in both poly- or mono-crystalline.\cite{4} 
Also, the present study is complementary to that for 
CeRhSb$_{1-x}$Sn$_{x}$, where a well defined Kondo-lattice insulating has been 
destroyed at critical value of $x\simeq 0.12$, where a quantum critical point 
and a phase transition KI$\rightarrow$NFL
have been detected. Here we did not observe such a quantum phase transition to 
the KI state, as the pure system CeNiSn exhibits already the KI semi-insulating 
behavior (cf. Fig. 1). Nevertheless, the stabilization of the full-gap state is an 
interesting phenomenon by itself, as the combined role of the disorder and the 
increased carrier concentration (and atomic disorder, to a lesser extent)
can be seen in a clear fashion.

\begin{figure}
\begin{flushleft}
\hspace{25mm}\includegraphics[width=0.4\textwidth]{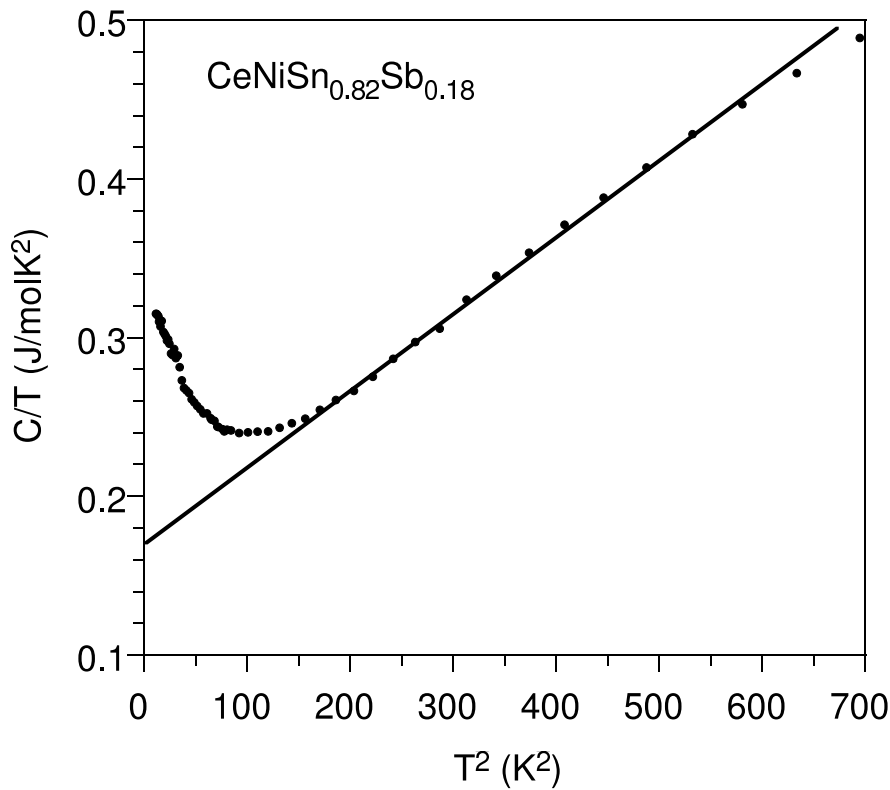}
\end{flushleft}
\caption{Temperature dependence of the (molar) specific heat $C/T$ - for 
CeNiSn$_{0.82}$Sb$_{0.18}$. The extrapolated value of the 
linear-specific-heat coefficient is 
$\gamma=174$~mJ/molK$^{2}$}
\end{figure}

One should note that a gradual evolution from the heavy fermion state ($T\gtrsim 
10$~K) into a Kondo semiconductor ($T\leq\Delta$) is related to the {\em onset
of localization\/} of $4f$ electrons with a concomitant formation of the 
compensation cloud bound to those localized moments. 
The effect is {\em collective\/} because {\em almost all\/}
Ce$^{3+}$ moments in the lattice are being compensated.
In the opposite
regime of high temperatures, the quantum coherence ({\em itineracy\/}) of 
$f$ electrons is destroyed by the thermal disorder. So, in principle,
we should distinguish between the coherence temperature $T_{coh}$
and the effective Kondo temperature $\Delta$; at least for the system evolving
with the decreased temperature from metal with localized $f$ electrons, through
the (moderately) heavy-fermion phase, to a Kondo semiconductor 
(insulator at $T=0$) or a semimetal. The heavy-fermion state 
rquiring itineracy of electrons,
takes place 
in the temperature range $\Delta\lesssim T\lesssim T_{coh}$.
The scaling $\chi\sim \rho^{-1}$ implies that there is a single energy scale
for both the thermal activation ($\rho$) and the singlet binding ($\chi$).

The semimetallic nature of CeNiSn means that the sizeable part of the
hybridization between $4f$ and the remaining valence electrons is of
intersite nature in that case and vanishes in some directions in reciprocal
space. With the Sn doping, the intraatomic part of this hybridization
dominates and leads to the reduction of the density of states at the
Fermi level, as seen in our photoemission data to be analyzed elsewhere.
In effect, the intraatomic Kondo coupling is also strengthened upon
Sn substitution.

In summary, we have shown that a Kondo semiconducting state can be 
created by the Sb doping of semimetallic CeNiSn. 
Let us stress again, atomic disorder introduced 
by the doping is not of primary relevance, but the extra electrons introduced with the 
Sb. The fractional (noninteger) number of introduced electrons must create a 
collective bound state of the Kondo type, and eliminates other possibilities, 
namely, the formation of either a Kondo band insulator or a Mott-Hubbard or Anderson 
insulating types of states. 
The formation of the collective Kondo-singlet state is determined 
by a (essential) reduction
of the magnetic susceptibility in low-$T$ range, as well by the presence of the scaling
$\chi(T)\rho(T)=const$, not only by the activated behavior of $\rho(T)$.
This work should be supported further by experiments 
on high-quality monocrystalline samples.

The authors acknowledge the financial support of the Ministry of Science and 
Higher Education, Grants Nos. 1 P03B 052 28 and P03B 001 29. J.S. thanks also 
the Polish Foundation of Science (FNP) for a senior fellowship for the years 
2003-7. The work has been performed under the auspices of the European Program 
COST P-16: {\em Emergent Behaviour in Correlated Matter\/}.
 The authors are also
grateful to Dr. Robert Podsiad\l{}y for technical help.


\begin{thebibliography}{99}

\bibitem{1}
S. K. Malik and D. T. Adroja, Phys. Rev. B {\bf 43}, 6277 (1991);
H. Kumigashira, T. Sato, T. Takahashi, S. Yoshii, and M. Kasaya,
Phys. Rev. Lett. {\bf 82}, 1943 (1999).

\bibitem{2}
Cf. e.g. T. Takabatake, F. Teshima, H. Fuji, S. Nishigori,
T. Suzuki, T. Fujita, Y. Yamaguchi, J. Sakuray, and D. Jackard, 
Phys. Rev. B {\bf 41}, 9607 (1990);
in the monocrystalline samples a plateau in temperature dependence
of $\rho(T)$ has been
observed, see: T. E. Mason, G. Aeppli, A. P. Ramirez, K. N. Clausen, 
C. Broholm, N. St\"{u}cheli, E. Bucher, and T. T. M. Paalstra,
Phys. Rev. Lett. {\bf 69}, 490 (1992).


\bibitem{3}
a) A. \'Slebarski and J. Spa\l{}ek, Phys. Rev. Lett., {\bf 95}, 046402 (2005);
b) J. Spa\l{}ek, A. \'Slebarski, J. Goraus, L. Spa\l{}ek, K. Tomala, 
A. Zarzycki, and A. Hackemer, Phys. Rev. B {\bf 72}, 155112 (2005).

\bibitem{4}
The Kondo insulator composed of a filled hybridized band of heavy quasiparticles 
has been considered in: S. Doniach and P. Fazekas, Adv. Phys. {\bf 65}, 1171 (1992);
R. Doradzi\'nski and J. Spa\l{}ek, Phys. Rev. B {\bf 56}, R14239 (1997); {\em ibid.\/} 
{\bf 58}, 3293 (1998). For a brief review see:
J. Spa\l{}ek and R. Doradzi\'nski, Acta Phys. Polonica A {\bf 96}, 677 (1999); {\em ibid.\/} {\bf 97}, 71 (2000).

\bibitem{5}
The effect of inclusions in CeNiSn has been discussed by Mason et al. Ref.~\onlinecite{2};
see also: G. Nakamoto, T. Takabatake, Y. Bando, F. Fuji, K. Izawa,
T. Suzuki, T. Fujita, A. Minami, I. Oguro, L. T. Tai, and A. A. Menovsky,
Physica B {\bf 206-207}, 840 (1995); G. Nakamoto, T. Takabatake, 
F. Fuji, A. Minami, K. Maezawa, I. Oguro, and A. A. Menovsky,
J. Phys. Soc. Japan {\bf 64}, 4834 (1995).

\bibitem{6}
A. \'Slebarski, J. Spa\l{}ek, M. Gam\.za, and A. Hackemer, 
Phys. Rev. B {\bf 73}, 205115 (2006).


\end{thebibliography}
\end{document}